\documentclass[twocolumn,superscriptaddress,10pt,aps,showpacs,reprint]{revtex4-1}

\usepackage{multirow}
\usepackage{amsmath}
\usepackage{amssymb}
\usepackage{microtype}
\usepackage{graphicx}
\usepackage{appendix}
\usepackage{color}

\begin{document}

\title{Efficiency and shrinking in evolving networks}

\author{Arianna Bottinelli}
\affiliation{NORDITA, Stockholm University, Roslagstullsbacken 23, SE-106 91, Stockholm, Sweden}

\author{Marco Gherardi}
\affiliation{Dipartimento di Fisica, Universita degli Studi di Milano, via Celoria 16, 20133 Milano, Italy}

\author{Marc Barthelemy}
\email{marc.barthelemy@ipht.fr}
\affiliation{Institut de Physique Th\'{e}orique, CEA, CNRS-URA 2306, F-91191, 
Gif-sur-Yvette, France}
\affiliation{CAMS (CNRS/EHESS) 54, Boulevard Raspail, 75006 Paris, France}

\begin{abstract}

  Characterizing the spatio-temporal evolution of networks is a central topic in many disciplines.  While network expansion has been studied thoroughly, less is known about how empirical networks behave when shrinking.  For transportation networks, this is especially relevant on account of their connection with the socio-economical substrate, and we focus here on the evolution of the French railway network from its birth in 1840 to 2000, in relation to the country's demographic dynamics.  The network evolved in parallel with technology (e.g., faster trains) and under strong constraints, such as preserving a good population coverage and balancing cost and efficiency. We show that the shrinking phase that started in 1930 decreased the total length of the network while preserving efficiency and population coverage: efficiency and robustness remain remarkably constant while the total length of the network shrinks by $50\%$ between 1930 and 2000, and the total travel time and time-diameter decreased by more than $75\%$ during the same period. Moreover, shrinking the network did not affect the overall accessibility, with an average travel time that decreases steadily since its formation. This evolution leads naturally to an increase of transportation multimodality  (such as a massive use of cars) and shows the importance of considering together transportation modes acting at different spatial scales. More generally, our results suggest that shrinking is not necessarily associated with a decay in performance and functions, but can be beneficial in terms of design goals and can be part of the natural evolution of an adaptive network.

\end{abstract}


\maketitle

\section{Introduction}

The evolution of networks has been the subject of numerous studies and books \cite{Barrat:2008,Latora:2017,Bottinelli:2017} and concerns
different fields, ranging from biology to transportation engineering \cite{Xie:2011,Tero:2010,Bottinelli:2015}. Many measures were defined and many models were proposed to describe the growth of these systems, but some important questions remain unanswered.

First, many networks interact with a substrate, and the question of the co-evolution of these components is still open. This interplay is especially relevant for transportation infrastructures, which are connected to the socio-economical conditions of the territory.  Indeed, these networks do not evolve in empty space, and the constraint of efficiency naturally imposes a coupling with the local population density. Railway networks are probably the best example of such a system, where the relation between network structure and the substrate is governed by complex feedbacks \cite{Xie:2011,Barthelemy:2018}. In the case of the French railway system, for instance, a recurrent debate revolves around the existence of a `structuring effect', whereby investments in transportation infrastructures have positive effects on productivity, demography, and the economy \cite{Mimeur:2018}. 

Second, almost all studies have been concerned with the expansion and growth of networks.  However, networks can evolve by alternating periods of increase and decrease of the number of nodes and links, and very little is known about this shrinking regime. This is true from a theoretical point of view (with the notable exception of a simple model proposed in \cite{Moore:2006}), but even more so from the point of view of empirical studies. Shrinking dynamics has been partially explored in the case of natural transport networks. For example, in laboratory conditions, the Argentine ant builds globally optimized transport networks that connect spatially separated nest~\cite{Latty:2011}. Such structures are achieved by initially creating several connections which later are either abandoned or amplified, causing the network to loose connections but not nodes. A similar pruning process is observed in the slime mould ({\it Physarum}), large multicellular organisms constituted by a network of tubes that circulates nutrients and signals~\cite{Tero:2010}. In both cases, the underlying mechanism is a self-organized positive feedback process where the passage of ants or nutrients reinforces pheromone traces~\cite{Perna:2012} or widens the slime's tubes~\cite{Ma:2013}.  While these are interesting examples, these networks probably evolve through very different processes compared to man-made infrastructures, such as roads, railways, or pipelines, where planning is often centralized. For transport infrastructures, the main design goal is to obtain a high transport capacity at a reasonable cost: cost and efficiency appear naturally as critical parameters governing the formation and evolution of these systems, sometimes at the expense of resilience \cite{Tero:2010,Bottinelli:2015,Bottinelli:2017}.  In the case of railway systems, in addition to the coupling with the population density, the network is also affected by technological advances that propose new and faster means which can be a cause of shrinking effects in these networks: older, slower lines can be abandoned as new faster lines appear, resulting in a global decrease of the total length of the network and its number of nodes. We thus apparently face here a trade-off problem: abandoning smaller lines and favoring faster lines by keeping at the same time a reasonable level of population coverage. This is a particular illustration of the competition between global social optimum and individual comfort \cite{Grauwin:2009,Bouchaud:2013}, and we could ask how the social optimum evolves during these various changes. More generally, one can ask how an evolutionary view could help to understand the development of transportation networks \cite{Bertolini:2007}, by considering them as far from equilibrium processes which behave in an evolutionary fashion, implying in particular that the focus in planning should be on enhancing the resilience and adaptability of these systems.

These two fundamental questions are particularly relevant for a range of physical systems, from spatial networks such as transportation infrastructures (power grid, etc), to other systems where nodes or links can disappear (e.g. in biology or in computer sciences). In this paper, we address these questions by empirically analyzing the evolution of the French railway network, from its birth in 1840 to 2000, in correlation with the evolution of French communes' population. We will characterize and discuss in detail both the growing and the shrinking phase, and how these phases fit in a larger picture of network evolution. This crucial example of a country-wide transportation network will also allow us to address the problem of shrinking networks and their coupling with the substrate structure. In particular, we will analyze the relationship between railway accessibility and population change as well as the changing spatial relations among national, regional, and local scales.

\section{ Evolution of the network}

\begin{figure}[h!]
\centering
\includegraphics[width=0.5\textwidth]{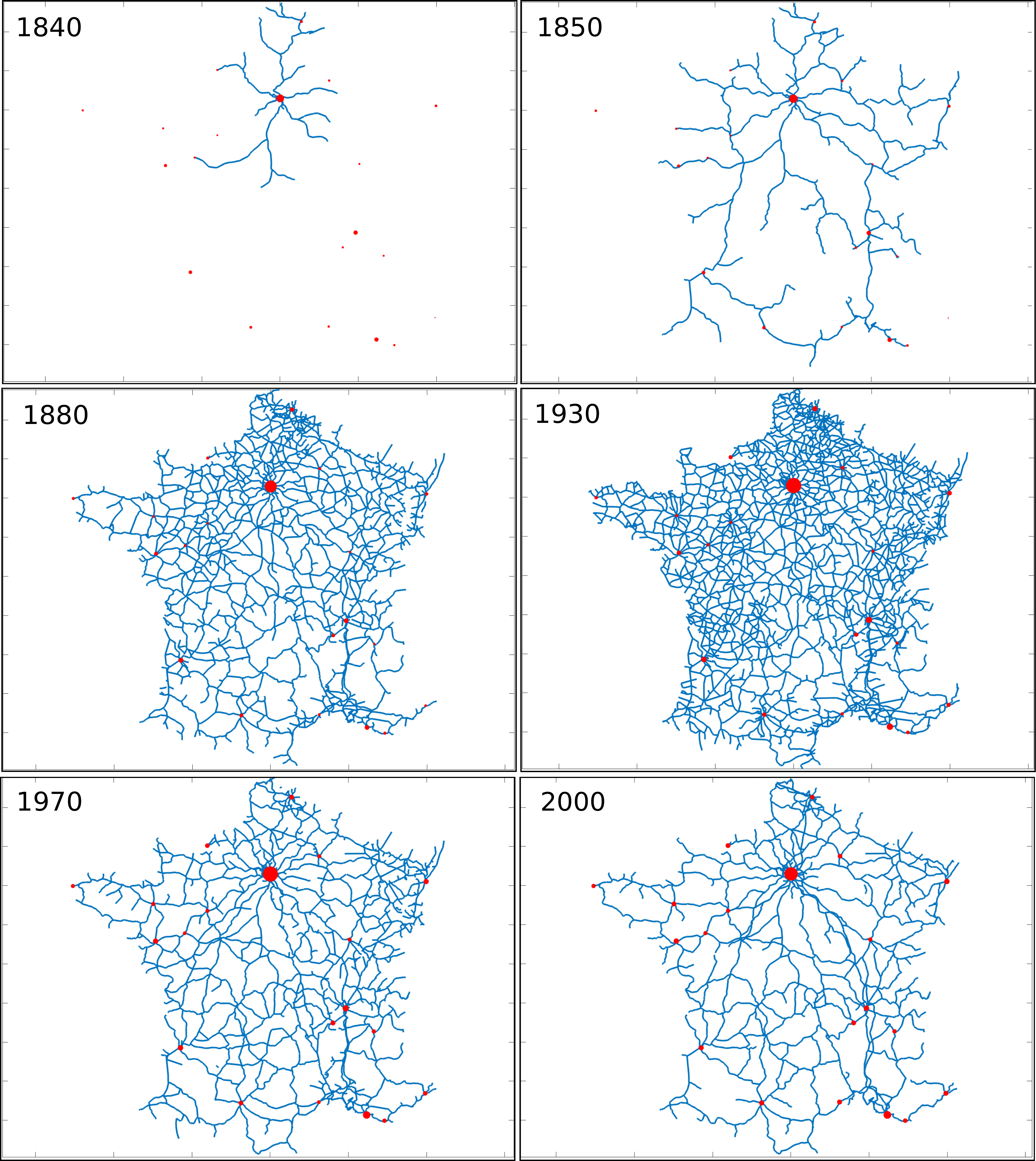}
\caption{\small Stages of the evolution of the railway network for selected years. Red dots are the 20 largest communes, the size of the dot is proportional to the commune's population size in the considered year. }
\label{fig:maps_netw}
\end{figure}

Between 1800 and 1900, the increasing industrialization caused a general trend in Europe of people moving from the countryside to cities. This urbanization process was relatively slow in France, and rural population remained the majority until 1930. Concerning the French railway network, different periods marked its evolution~\cite{Thevenin:2016,Mimeur:2018} (Fig.~\ref{fig:maps_netw}): (i) first, between 1830 and 1860, the government started a National railway policy and assigned to 6 monopoles the construction of 6 radial lines departing from Paris in order to reinforce the capital's centrality through connections with important cities. These monopoles were private companies that did not interact or connect with each other. (ii) The second period (1860-1890) witnessed the creation of more lines between Paris and other regional cities, and the creation of `lateral' lines connecting the initial 6 radial lines with each other. (iii) The third period (1890-1930), was mainly devoted to the creation of `lines of local interest', with the main goal of using the railway system to connect and modernize smaller towns and rural areas. Thus, the network reached its maximal expansion in 1920, while in the fourth (iv) period (1930-1950) the network underwent a contraction due to the modernization of the equipment and the elimination of local `narrow gauge lines', substituted by roads. (v) Finally, in the modern period (1980-2000), we assist to the creation of high speed lines (TGV) further reinforcing the use of main lines and the abandon of smaller local lines.

In this study, we will use two different datasets: one concerns the evolution of the French railway network, and the other contains the historical records of the French communes population. Railway network data are constituted by stations (nodes), with their geographic position, and rail track segments (links) characterised by their length, travel time, etc (for details about these datasets see the SI). More precisely, the network is composed by $N = N_T + S$ nodes, where $S$ is the number of stations and $N_T$ is the number of topological nodes (i.e. nodes that are not stations but are needed to indicate junctions or ramifications in the tracks). While for stations and track segments we have the explicit opening and closing dates, the total number of nodes $N$ and the number of stations $S$ at a certain time $t$ ($N_T$ is thus obtained by subtraction). This information is available starting from 1840 every 10 years, allowing us to reconstruct the railway system and to analyze it (see Fig.~\ref{fig:maps_netw}).

\begin{figure}[h!]
\centering
\includegraphics[width=0.5\textwidth]{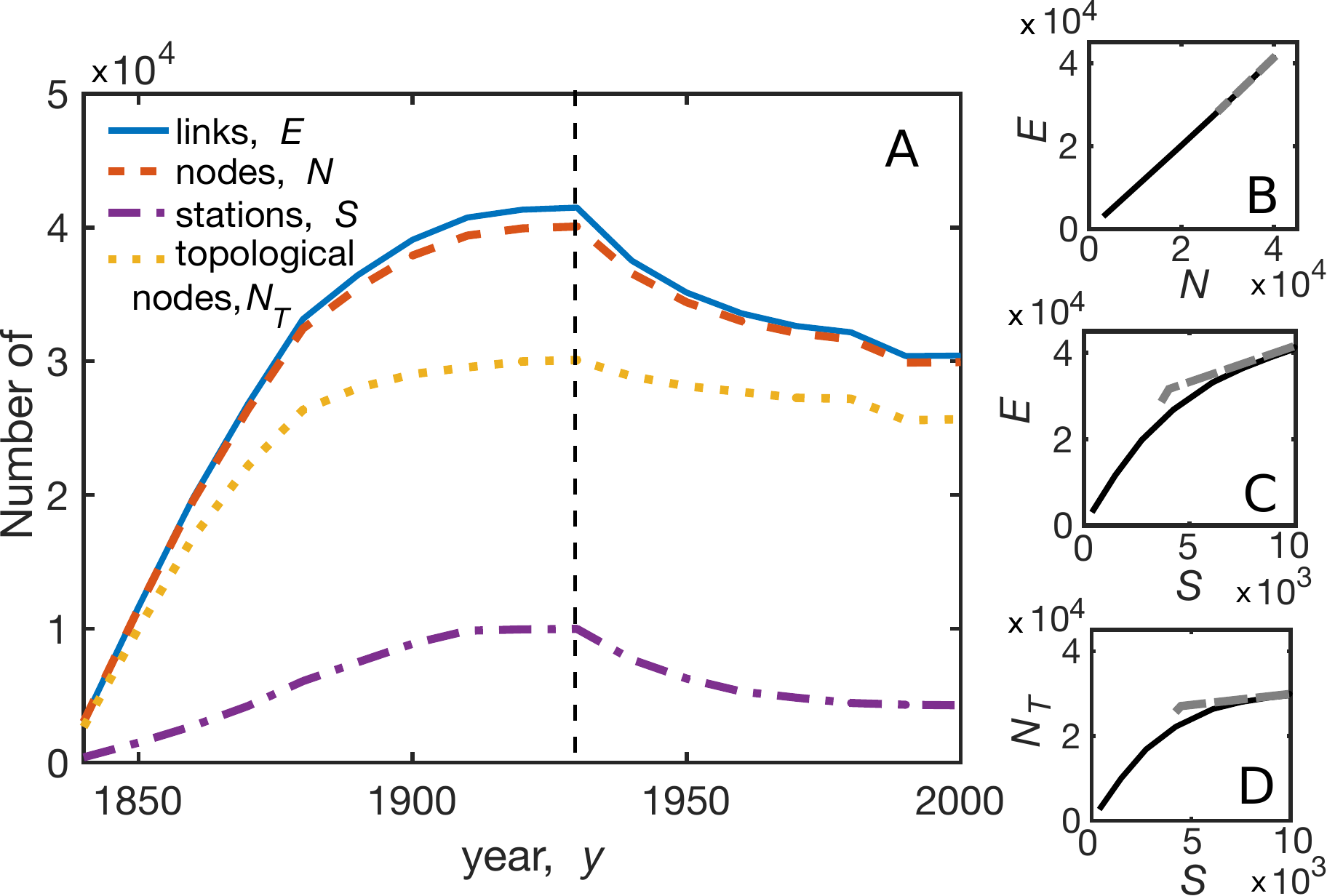}
\caption{\small  (A) Number of links, nodes, topological nodes, and stations constituting the railway network versus time. The vertical dashed line indicates the beginning of the shrinking phase in 1930. Number of (B) links as a function of nodes. Number of (C) links, and (D) nodes as a function of the number of stations. Full lines represent the growth phase, dashed lines the decreasing phase.}
\label{fig:links_nodes_time}
\end{figure}

As discussed also in \cite{Thevenin:2016}, we observe that the growth of the number of links and nodes slows down around 1880, increases until 1930, and then decreases [Fig.~\ref{fig:links_nodes_time} A]. 
The number of total nodes and links display a very similar temporal behavior suggesting that, especially before 1930, the growth rule was to add a node and a link at a time. The number of nodes versus the number of links displays a linear behavior which corresponds, as expected, to an average degree of order 2 (Fig.~\ref{fig:links_nodes_time} B). Taking $S$ as a reference allows us to clearly distinguish the two phases of growth and decrease of the network (Fig.~\ref{fig:links_nodes_time} C, D). In particular, $N_T$ and the total number of links $E$ grow faster than $S$, and the decrease seems mostly linear in both cases. 

\begin{figure}[h!]
\centering
\includegraphics[width=0.5\textwidth]{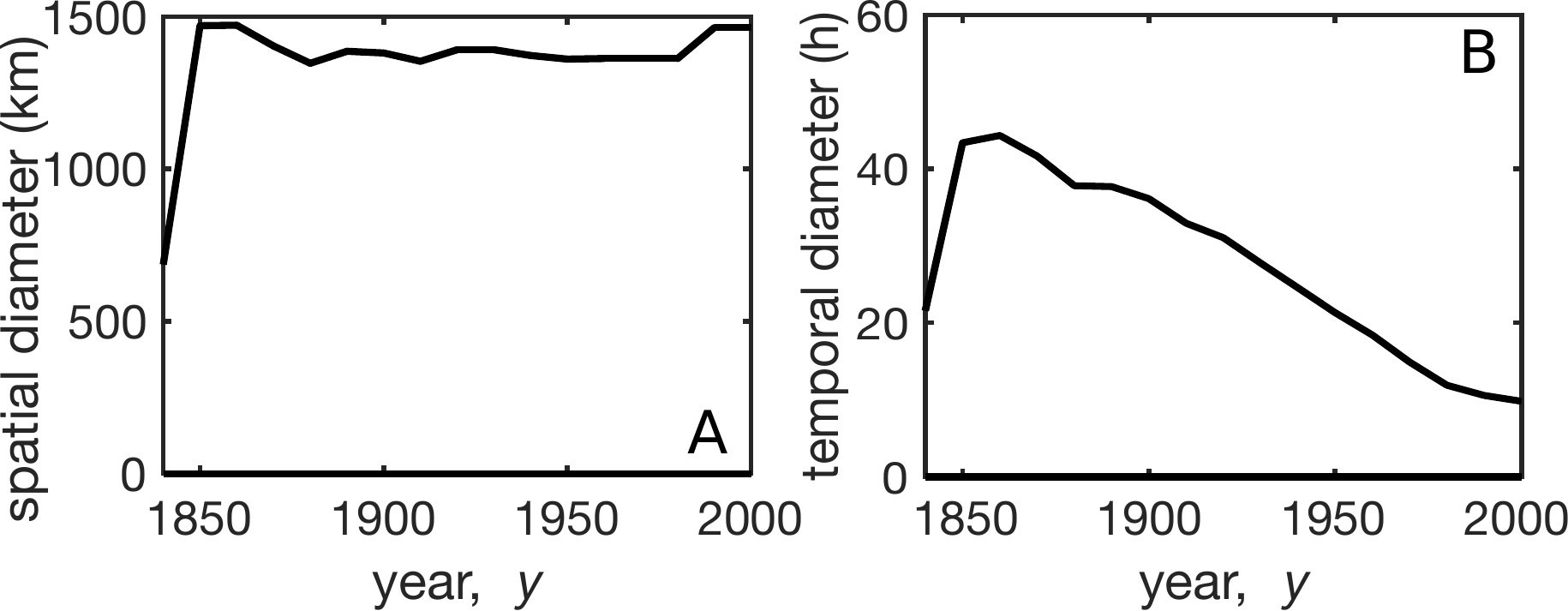}
\caption{\small  Diameter versus years computed for (A) the shortest path (in kms) and (B) the quickest path (in hours). }
\label{fig:msp}
\end{figure}

Another macroscopic measure that characterizes this network is its diameter, defined as the length of the longest shortest path between two points (see for example \cite{Clark:1991}). For a transport network, we can compute shortest paths in terms of distance with length of the tracks or travel time (so that the shortest path is the quickest path, measured in hours). We represent the evolution of the diameter for these two choices in Fig.~\ref{fig:msp}.

We observe that, after an initial quick growth, the `spatial diameter' is constant and of order the maximum size of the country ($\sim 1500$kms). In contrast, the `temporal diameter', based on quickest paths, displays a remarkable monotonous decrease. This is the first sign that the shrinking of the network is compensated in some way by technological advances (the increase in the speed of trains in this particular case). Other topological measures, such as the number of nodes of a given degree and the cyclomatic number, are reported in the SI (see Figs.~S2 and S3).

\section{ Cost, efficiency, robustness.}

\begin{figure}[h!]
\centering
\includegraphics[width=0.5\textwidth]{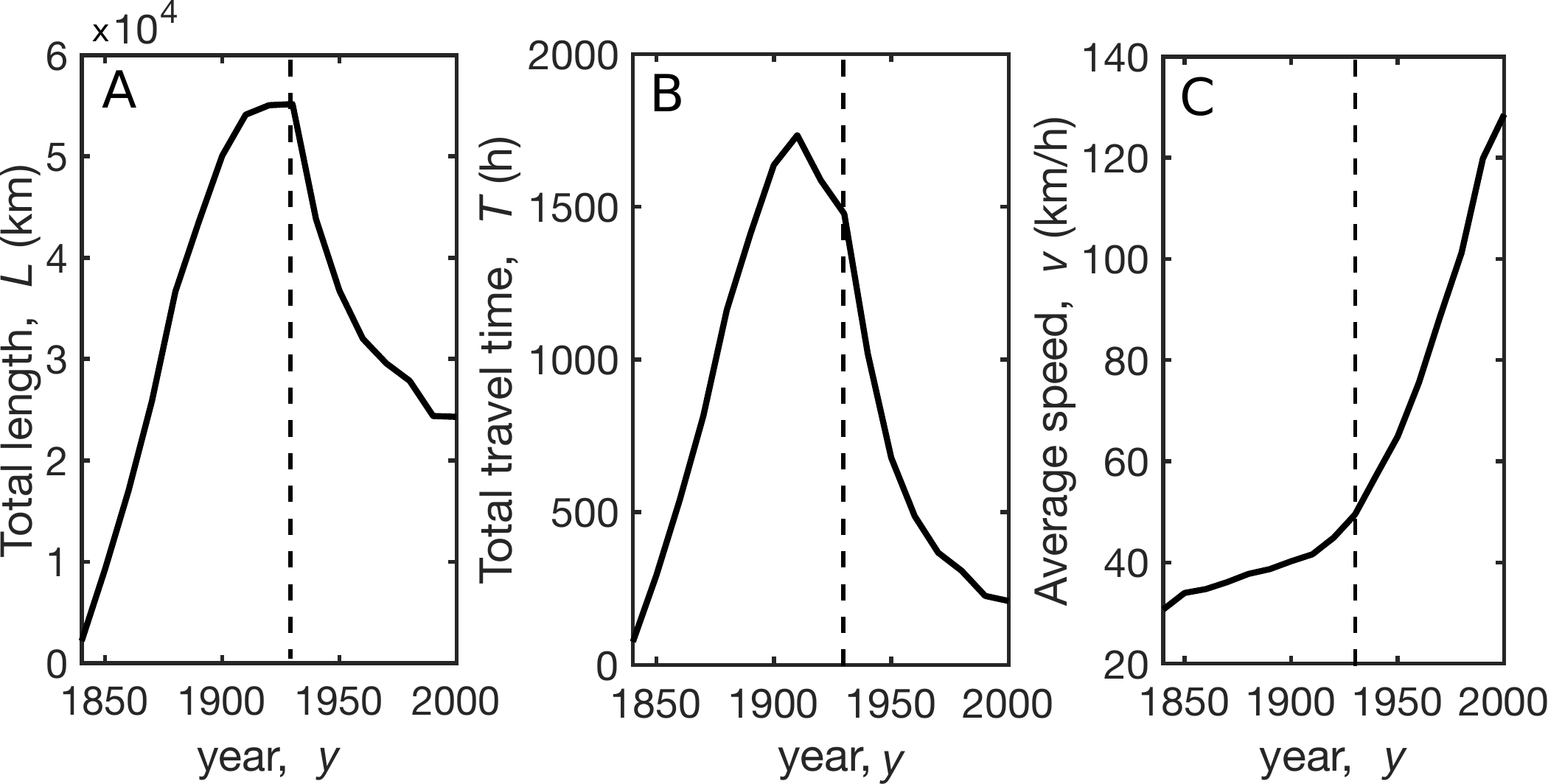}
\caption{\small Evolution of the (A) total length of the network, (B) total travel time, (C) average trains' speed. The vertical dashed line indicates the beginning of the shrinking phase in 1930.}
\label{fig:length_time_speed}
\end{figure}

A known challenge when designing transport networks is to balance between the network's cost, efficiency, and robustness~\cite{Latora:2007,Tero:2010,Xie:2011}. The cost of a spatial network $G$ built on $N$ nodes is usually estimated by its total length
\begin{align}
L(G) =\sum_{e \in E(G)} \ell_{e}
\end{align}
where $\ell_{e}$ is the Euclidean length of a link $e$ belonging to the set of links of $G$. We can also compute the total travel time $T(G)$ on the network (given by the sum of all travel times over all rail segments), or the average speed $V(G)=\sum_ev(e)/E$ where $v(e)$ is the speed on link $e$.  We plot 
these quantities versus years in Fig.~\ref{fig:length_time_speed}. 
These figures demonstrate that both the total travel time and the total length have a peak at the beginning of 1900, and then decrease until nowadays. For the total length, the decrease is mainly due to the elimination of `local narrow gauge lines', replaced by roads after 1930 (Fig.~\ref{fig:length_time_speed}(A)). The decrease in the total travel time starts slightly earlier due to the modernization of locomotives and to the systematic electrification of railway lines after 1920, and is further enhanced by the elimination of slow lines after 1930 and by the construction of TGV lines in 1980 (Fig.~\ref{fig:length_time_speed}(B)). These technological advances are well summarized by the evolution of the average speed (Fig.~\ref{fig:length_time_speed}(C)) which displays a constant increase, in particular after 1930.

\begin{figure}[h!]
\centering
\includegraphics[width=0.5\textwidth]{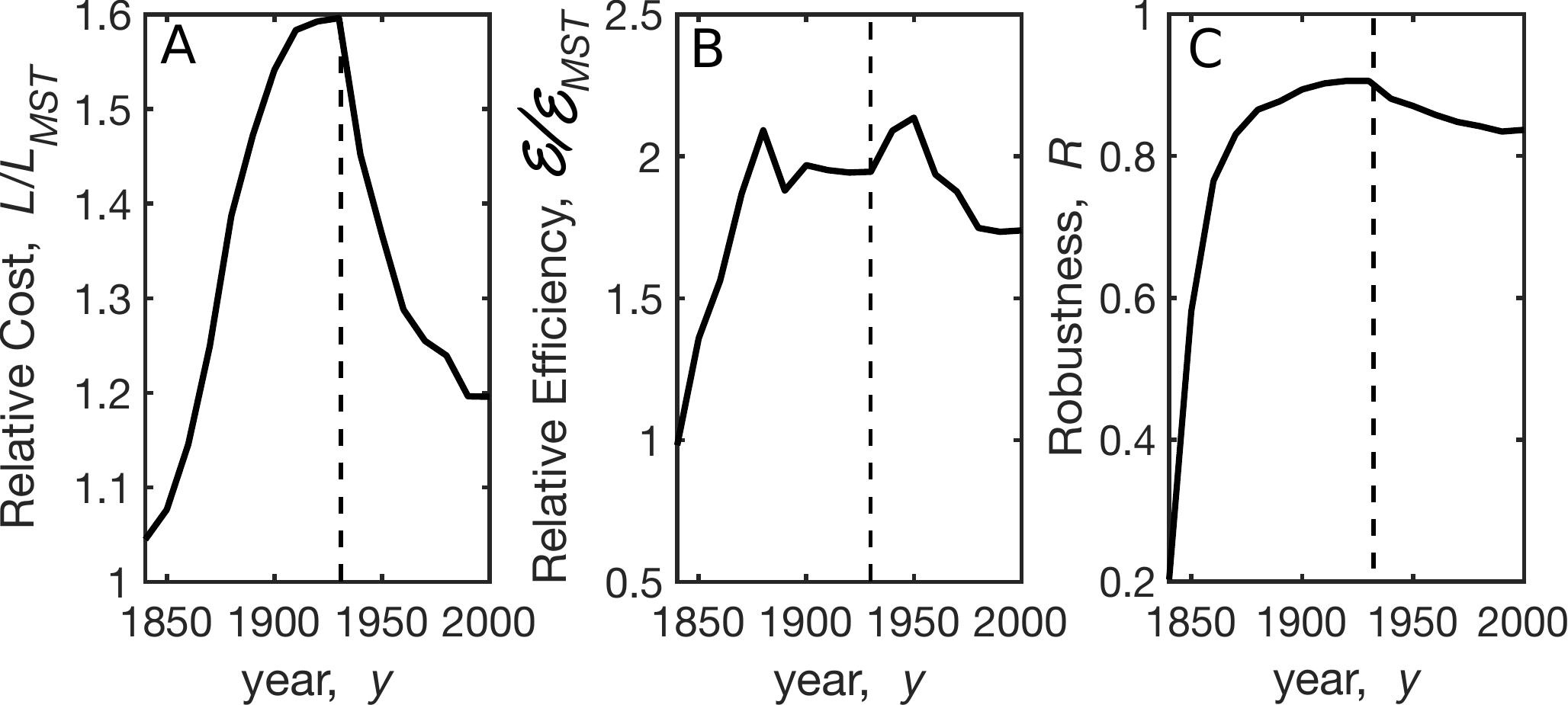}
\caption{\small  Evolution of the (A) Cost of the train network divided by the cost of the MST (constructed over the same set of nodes), (B) Efficiency of the train network divided by the efficiency of the MST, (C) Robustness. The vertical dashed line indicates the beginning of the shrinking phase in 1930.}
\label{fig:eff_cost_rob}
\end{figure}

In order to understand the order of magnitude of the network's cost, expressed by the total length, we can compare it to the most economical network that connects all the stations and topological nodes present at a certain time. This is the minimum spanning tree (MST, see for example \cite{Graham:1985}), which represents an excellent benchmark for spatial networks (see for example \cite{Tero:2010,Cardillo:2006} and \cite{Barthelemy:2018} and references therein). We can then construct the relative cost $L/L_{MST}$ for connecting the same set of nodes and see how this ratio varies with time (Fig.~\ref{fig:eff_cost_rob}(A)). We observe that the relative cost has a peak around 1930, indicating that the total length of the actual railway network is 1.6 times the minimum length needed to connect all stations. After 1930, this ratio decreases to 1.2 in 2000, showing the large reduction of costs that governed this period. 

While the total length is generally accepted as a good proxy for the cost of the network, several definitions of transport efficiency can be found in the literature~\cite{Latora:2001,Cook:2014,Aldous:2010,Barthelemy:2018}.  Efficiency is often regarded as one of the main design goals in planning and building transportation infrastructures~\cite{Gastner:2006,Bebber:2007}. Here, we follow~\cite{Latora:2001} and define it as \begin{align}
 \mathcal{E}(G) = \frac{1}{N(N-1)} \sum_{i \neq j \in G} \frac{d_E(i,j)}{d_N(i,j)} ,\label{eq:eff}
\end{align} 
where i and j are nodes in $G$, $d_E(i,j)$ is their Euclidean distance, and $d_N(i,j)$ is the length of the shortest path connecting them on the network $G$. 
With this definition, efficiency takes values between 0 and 1, and quantifies, on average, 
how much the shortest paths on the network are close to straight lines. We compare it to its value for the MST which is explicitly built by prioritizing cost reduction over efficiency, and we observe that, despite the strong reduction in nodes and links, the French railway network remains twice as efficient as the cheapest network connecting all the nodes (Fig.~\ref{fig:eff_cost_rob}(B), efficiency alone is shown in Fig. S4). We note that this measure does not take into account the speed of trains, but simply the ability of the network to transport passengers between communes along the straightest possible path. Another relevant measure for transport networks is robustness, or fault tolerance, computed as the probability of {\it not} disconnecting the network by removal of a random link~\cite{Tero:2010}. In the literature, robustness is used to measure the resilience of the network against the breakdown of its components (links in this case), and is remarkably high here (see  Fig.~\ref{fig:eff_cost_rob}(C); we recall that the MST is a tree thus its robustness is zero). 

During the radial and the capillarization expansion phases of the network, we observe a large increase of all these quantities. It is remarkable to observe how the decreasing phase after 1930 has greatly reduced costs while keeping efficiency and robustness almost unchanged. This is a probably consequence of the strong centralization of the national railway system, where the addition and deletion of lines, and thus network optimality, were planned ahead at the governmental level.  We also note that one can visualize the evolution of the trade-off between these design goals by plotting these quantities one against each other as in \cite{Tero:2010} which supports the fact that during the shrinking phase costs are reduced costs but the robustness and efficiency are almost unchanged (see Fig.~S4).

\begin{figure}[h!]
\centering
\includegraphics[width=0.5\textwidth]{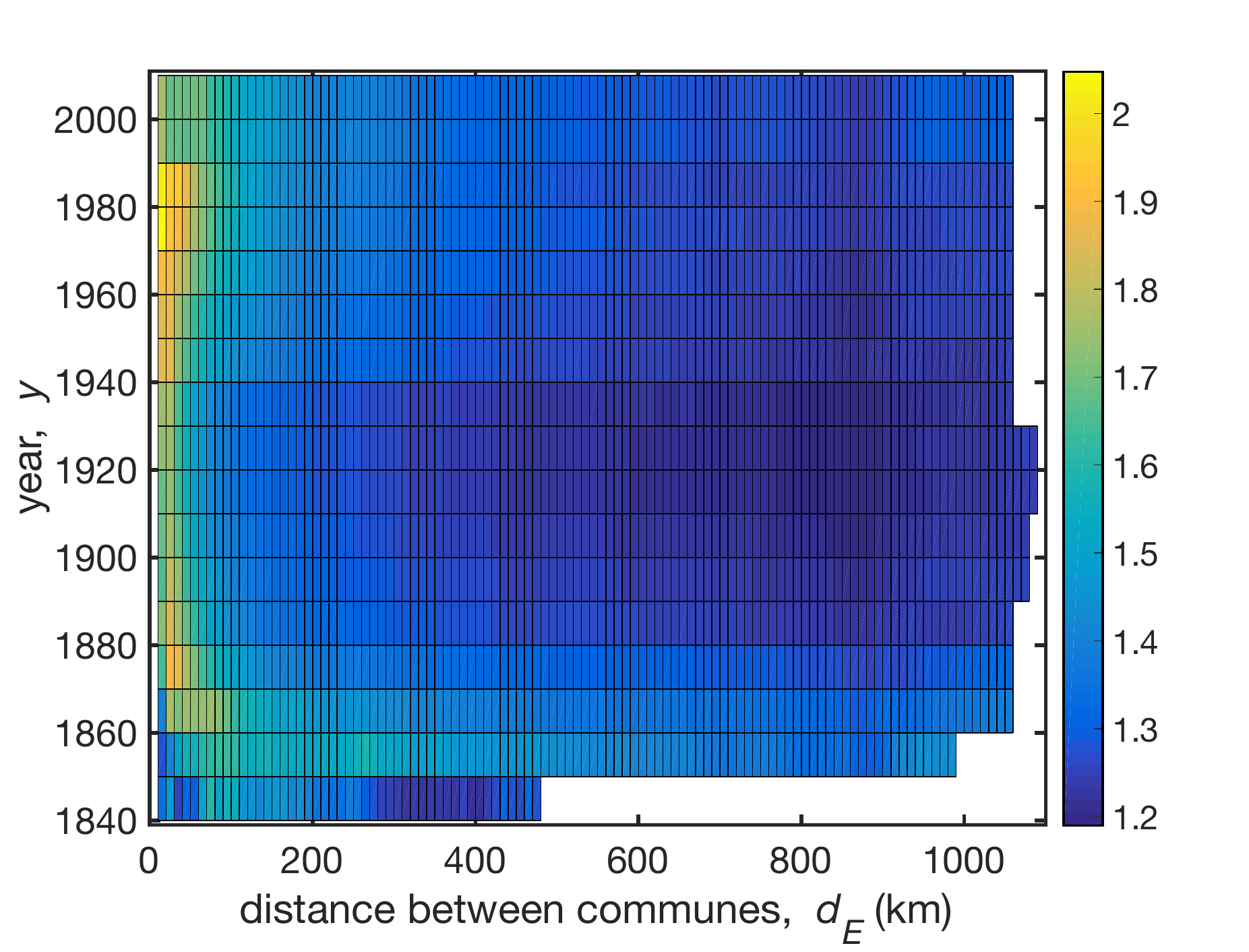}
\caption{\small Heat map showing the evolution of the detour index at different euclidean distances between communes (brighter is high detour index, darker is low detour index).}
\label{fig:detour}
\end{figure}

Another indicator (non-trivially) related to efficiency and which allows a clear characterization of the structural changes at various scales is given by the detour profile  $\phi$ \cite{Aldous:2010} which is defined as follows
\begin{align}
\phi(d)=\frac{1}{N(d)}\sum_{i,j\;s.t.\; d_E(i,j)=d}\frac{d_N(i,j)}{d_E(i,j)}
\end{align} 
and is a function of the distance $d$ and where $i$ and $j$ are communes with a station, $d_E$ is their euclidean distance, $d_N$ is their distance on the network, and $N(d)$ is the number of pairs of communes that are at distance $d_E(i,j)=d$. $\phi$ is larger than $1$ and indicates the average deviation from a straight line needed to travel on the transport network between any two communes at distance $d$. This measure is suitable for understanding the focus distance of the operations on the transport networks (expansion, pruning) through time. Moreover, it highlights which distances between communes were typically favored during the different phases of evolution of the network. For the French railway system, we observe a strong decrease of the detour profile at large distances, due to the construction of the main radial lines after 1860, which remains constantly low through network evolution (Fig.\ref{fig:detour}). This also has the effect of reducing the detour at shorter distances ($< 200$ km), although a peak remains at short-intermediate distances of order 20-100 km. This is the range of distances that was targeted in the following capillarization phase, which implied a strong reduction of the detour index above 30 km, but not below, as probably this was a reasonable distance to cover by walking or riding. The pruning phase starting in 1930 determines a gradual but significant increase in the detour index in the same distance range ($< 100$ km), that seems to correlate with the increasing speed of other transport means. For example, in 1980, it was already possible to cover 100kms by car in about an hour. We also observe that pruning at a local scale also slightly affects intermediate distances (100--600 km). Finally, the detour profile can be averaged over distances to monitor the time evolution of the network (see Fig. S5 and SI), obtaining a measure that is complementary to efficiency (Eqn.~\ref{eq:eff} and Fig.~\ref{fig:eff_cost_rob}(B)). 

\section{ Evolution of the population and coverage properties}

\subsection{Population and the network}

\begin{figure}[h!]
\centering
\includegraphics[width=0.5\textwidth]{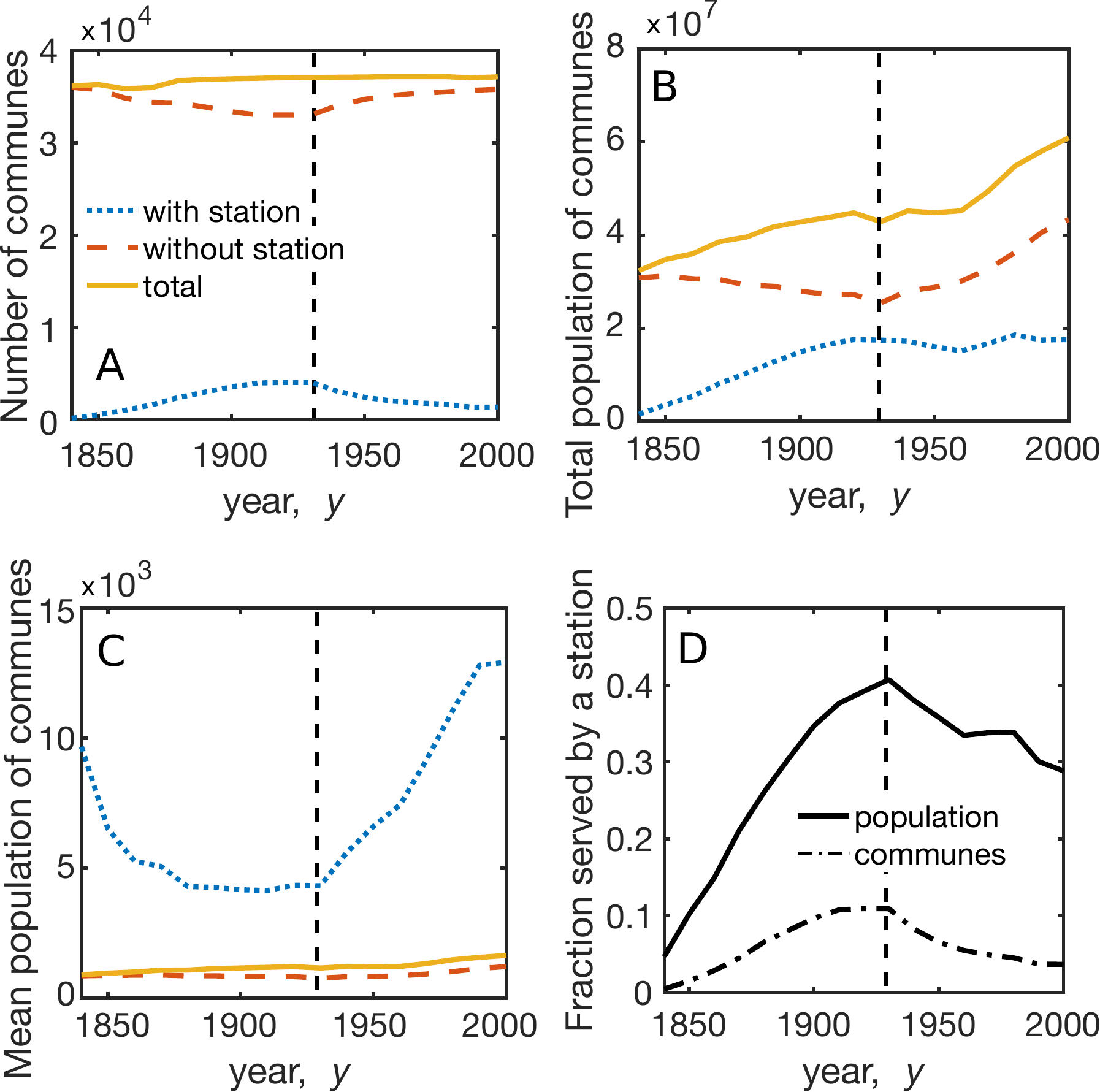}
\caption{\small Evolution of the (A) number of communes, (B) total population, (C) mean population, (D) fraction of communes and population directly served by a station. The vertical dashed line indicates the beginning of the shrinking phase in 1930.}
\label{fig:cs}
\end{figure}

The railway network co-evolves with the population distribution, and it is therefore important to characterize quantitatively the correlation between the network's extension and the population density at both a global and a local level. First, we consider the evolution of the number
of communes with and without a station (Fig.~\ref{fig:cs}(A)). As expected, we observe a peak around 1920 for the number of communes with a station, while the total number of communes is roughly constant. It is interesting to observe that, while the total population grows constantly, during the expansion phase of the network the growth of the population appears to occur mainly in communes with a station, while it is concentrated in communes without a station during the shrinking phase (Figure~\ref{fig:cs}(B)). Although this is expected in the growing phase it is a surprise to observe that the majority of the population is growing in communes without a station. This is consistent with the fact that the average population of communes with a train station displays a minimum around 1930, when many small communes were directly connected to the network as a result of the government's policy of reaching small countryside towns 
(Figure~\ref{fig:cs}(C); for the full distributions of population sizes see Fig.~S6).
After 1930, we then observe an increase of the average size of communes connected to the network. 
Overall, it is interesting to observe that, while the fraction of communes with a station is always lower than 0.1, in the moment of maximum expansion of the network almost half of the French population lived in a commune served by a station (Figure~\ref{fig:cs}(D)).

\subsection{Accessibility}

An important aspect of the relation between a transport network and its substrate is the network's accessibility. Several quantitative ways of estimating accessibility have been proposed in the literature, and the different approaches are reviewed in \cite{Handy:1997,Geurs:2004}. For instance in our case, a rough but straightforward way to estimate the railway's accessibility is to measure its {\it pedestrian accessibility}~\cite{Thevenin:2016}, defined as the Euclidean distance $d_E$ between a commune $c$ to its nearest train station $s$: 
$A_c=  d_E(c,s)$ (for a commune $s'$ that has a station we then have $A_{s'}=d_E(s',s')=0$). By averaging over all communes (with or without a train station) and by weighting with a commune's population $P_c$, we obtain the network's average pedestrian accessibility
\begin{align}
\label{eq:accessibility}
\langle A\rangle=\frac{\sum_c P_c \, A_c}{\sum_c P_c}.
\end{align}
This quantity depends on how the network extends in the territory with respect to the local population density, and measures the typical distance needed by a random individual to reach the nearest station. 
Note that the definition [\ref{eq:accessibility}] implies that the larger the railway network coverage and the lower its accessibility.
\begin{figure}[h!]
\includegraphics[width=0.5\textwidth]{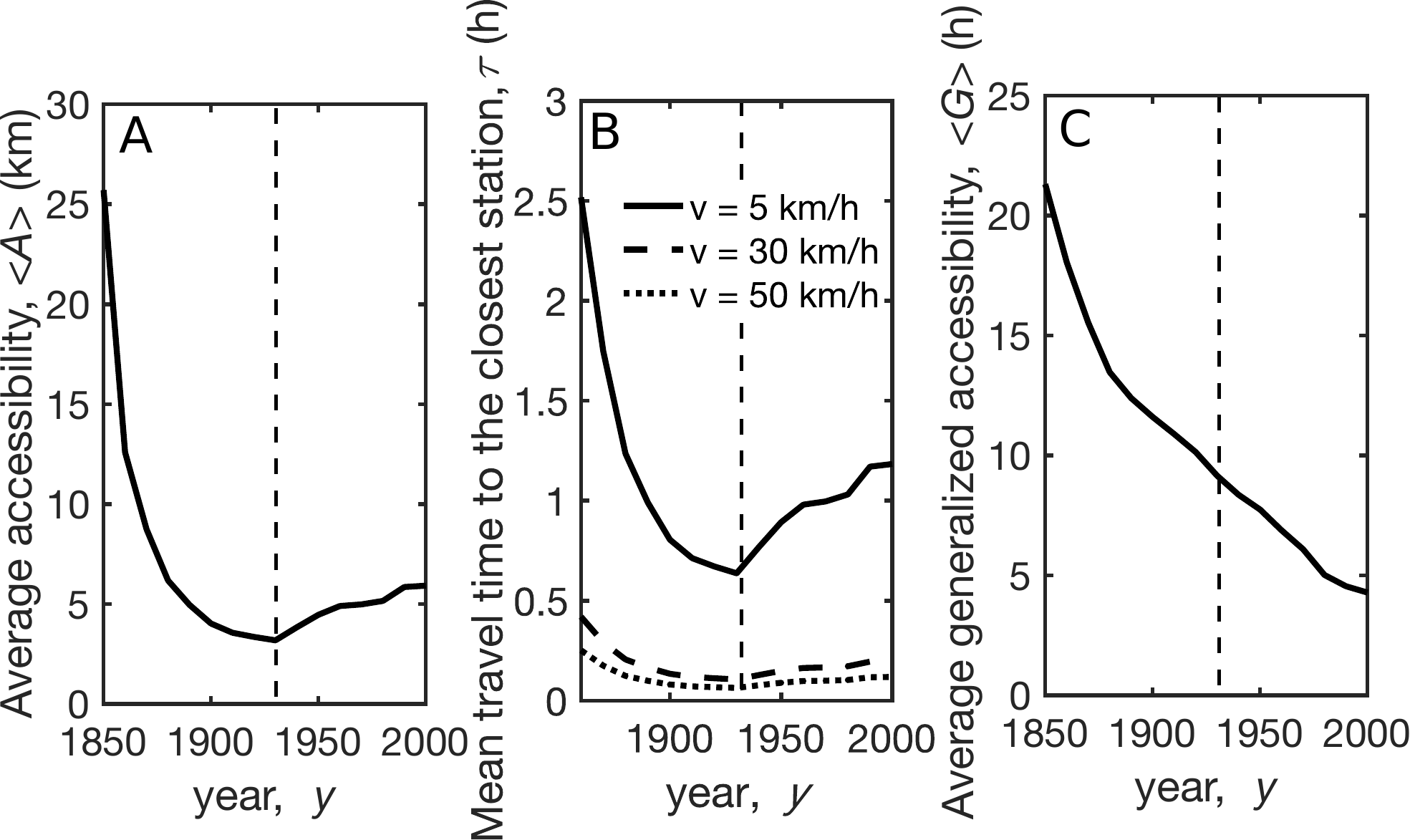}
\caption{\small  (A) Average pedestrian accessibility; (B) Average travel time to reach the closest station for a random French citizen at different values of the transport velocity $v_p$. (C) Average generalized accessibility versus time computed with $v_p = 5$ km/h. Data are shown starting from 1850 to better appreciate the variations in recent times ($\langle A \rangle (1840) = 162$ km, $\langle \tau \rangle (1840) = 32$ h). The vertical dashed line indicates the beginning of the shrinking phase in 1930.}
\label{fig:newaccess}
\end{figure}
Figure ~\ref{fig:newaccess}(A) shows that from the birth of the railway network to the moment of maximum expansion and capillarization in 1930, the average travel distance per person to the closest train station dropped from 25 km to less than 5 km. After 1930, the removal of the smallest lines increased this average distance which however remained rather contained (slightly above 5 km).

Another way to look at accessibility is through the average travel time required to reach the closest train station $\langle\tau\rangle=\langle A\rangle/v_p$, where $v_p$ is the typical speed of complementary transport means. At the beginning of the century, we can assume the main transport mode was walking ($v_p \approx 5$ km/h), and that, later on, coaches ($v_p \approx 30$ km/h) and cars ($v_p \approx 50$ km/h), together with better road infrastructures, increased this velocity. In Figure~\ref{fig:newaccess}(B) we compare the average travel time for different transportation speeds. We observe that, due to the capillarization of the network in the territory, from 1900 to 1950 it was possible, on average, to walk to the closest station within one hour. While this may have been a completely reasonable option at the beginning of the century, the current lifestyle and needs require to take either public transportation, when available, or to use the car, reducing the average travel time to about 10-15 minutes. By keeping into account the average road transport speed typical of each decade, we observe a decrease of $\langle\tau\rangle$, despite the decreasing amount of communes being directly served by a station (as shown in Fig.~\ref{fig:cs}A and D). Therefore, even if the network is shrinking after 1930, the stations that remain open seem to have been chosen strategically such that the pedestrian accessibility is kept roughly constant, in particular if we take into account the possibility of other transport modes. 

So far, we have characterized how effectively the transport network extends in the territory with respect to the spatial distribution of communes and population. In order to quantitatively assess how well it connects two random communes in the French territory we also need to take into account the average travel time on the network. We thus weight each link in the network with its travel time, which, in contrast with the track's length, changes in time with technological improvements. This leads us to define the generalized accessibility $G_c$ as
\begin{align}
G_c = v_p^{-1} A_c + \frac{1}{S-1}\sum_{s'} t_N(s,s'),
\end{align}
where $t_N$ is the shortest time-weighted path between $c$'s closest station ($s$) and any other station $s'$ ($S$ is the total number of stations at time $t$). $G_c$ is thus the average time needed to travel from a commune $c$ to any other point in the network. We then average over all communes weighted with their population size, and obtain the average generalized accessibility $\langle G\rangle=\sum_cP_cG_c/P$ where $P$ is the total population in France (at a given time) which represents the average time needed by 
French citizens to reach the railway network and travel to any other station. 
Figure~\ref{fig:newaccess}(C) shows that $\langle G \rangle$ computed with fixed $v_p = 5$ km/h decreases from 22 hours in 1850 to less than 5 hours in the year 2000 (see also Fig.~S7 and S8 for the geographical distribution of the accessibility measures.)

\section{Discussion}

The shrinking of the network characterized by a decrease of the number of stations and lines started in the 1930s, and was reinforced in the 1980s by the appearance of high speed trains, that led to a further trimming of the network's smallest and slowest lines. Removing links and nodes from the network may affect negatively the general transport performances of the railway, potentially reducing the efficiency of the network and increasing the travel times for a large sector of the population. However, we find that efficiency indicators are not negatively affected by the country wide re-organization of the railway network. At a topological level, efficiency and robustness remain remarkably constant while the total length of the network shrinks by 50$\%$ between 1930 and 2000. At an efficiency level, thanks to technological improvements, the total travel time and time-diameter decreased by more than 75$\%$ during the same period. Moreover, shrinking the network did not affect the overall accessibility, when considering the distribution of population across the territory. Indeed, the average travel time decreased steadily since its formation. All these results seem to point to one conclusion: even if pruning the network and closing stations and lines may initially appear as purely cost-driven governance, it seems that this evolution is rather natural and beneficial in terms of design goals. In contrast with naive intuition, taking advantage of new technologies in both railway and road transportation further improved the average network performances for covering the territory.

Our analysis shows the importance of considering the evolution of transportation infrastructure 
in conjunction with the socio-technological substrate and technological improvement. The increasing quality of roads and mass availability of cars decreased the access time to train stations and favored the reorganization of the French railway system. In this sense, removing smaller local lines was concomitant with an increase of multimodality in the transportation system. With an eye to the current debate on global warming and sustainable transportation, it seems necessary to scrutinize decisions such as substituting local electrified train lines with roads. Overall, our quantitative analysis suggests that the French railway system provides an efficient and sustainable large-scale transport infrastructure, which could be better exploited by strategically planning other public transportation means, acting at a smaller spatial scale. Relatively slower, but collective, transport means (e.g. electric or biogas buses) could provide a better trade-off between transportation efficiency and environmental impact. At a more fundamental level, our results promote a unified framework where network and substrate evolution are considered jointly, and where mutual influences are taken into account.  In the case studied here, our measures suggest that the transformation of the French railway during the last two centuries is associated to a profound scale-dependent transport mode diversification, and that shrinking is not necessarily associated with a decrease of efficiency but can be a part of the natural co-evolution of this system.


\vspace{0.5cm}
{\bf Acknowledgements}
We thank Thomas Thevenin for the train network data. We also thank Maurizio Gribaudi and
the Geohistoricaldata group for the population data.

\section*{Supplementary Material}

\subsection{Data}

\subsubsection{The network}

\begin{figure}
\centering
\includegraphics[width=0.4\textwidth]{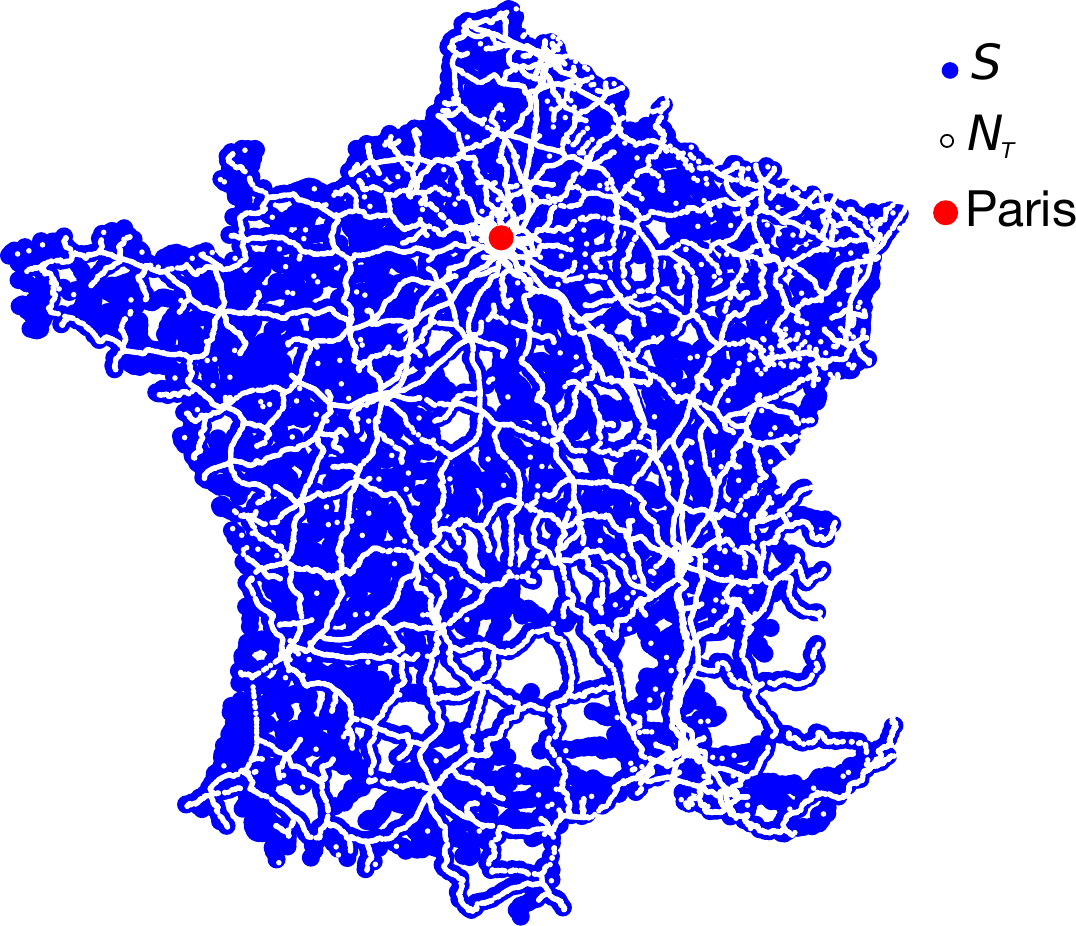}
\caption{\small Spatial plot of all available stations (blue circles) and topological nodes (white dots). Paris is indicated by the large red circle. }
\label{fig:maps}
\end{figure}

Data are composed by a set of Nodes $N$ and Arcs $E$ (edges). Nodes can be actual stations $S$ (for goods and/or people), or ``topological'' nodes $N_T$ where lines cross or change direction (Fig.~\ref{fig:maps}). The available information about all nodes are:
\begin{enumerate}
  \item thier identification number ID,
  \item the $X$ and $Y$ coordinates.
\end{enumerate}
Additionally, for each station we know its nature (passengers, goods, both), the dates when it was opened and closed, the INSEE number (ZIP code), name and status of the corresponding commune, and the region the commune is in. For each INSEE the population $P(t)$ of the commune is available every 10 years from 1840 to 2000.

Each arc $e$ is determined by the ID number of the initial and final nodes, the dates when it was opened and closed, the associated length of the arc, and the travel time every 10 years from 1840 to 2000. 

Note that in Arcs there were 72 couples of duplicated edges, i.e. links
that connected the same couple of nodes, but had different lengths and
travelling times, and sometimes opening and closing dates. This might be
due to the presence of more then one line connecting two
nodes, but sometimes the lengths of the two matching arches are quite
different. As an example, there are two arcs connecting nodes 291 and
295, the first is long 318.5 m, the second 1607.3 m, and travelling
times are also different. When this occurred, we keept the link with the longest opening
span and eliminated the other.  Since the amount of such duplicated links is relatively small, compared to the total number of links, we do not expect that 
eliminating them affects the large-scale properties of the
network.

\subsubsection{Population data}

The original database ``BDCassini'' provided us with population data from 1840 to 2000 for about 37,192 communes (not all communes existed at the same time). 10,187 of them had a functioning station some time during the considered dates. 
For each commune we also have information about when it was born and possibly when it was abandoned, and the opening and closing dates of the station if they had one.

Multiple stations with the same INSEE are present in the dataset,
as one city may have more than one station.
When this occurs, we collapse all these stations to a single node in the network,
whose population corresponds to the commune with that INSEE.
In the special case of Paris, several stations belonging to the city have different INSEE
numbers (most of them begin with 751).
We keep the coordinates, INSEE, and
population data of CHATELET-LES HALLES and assign it to all the
stations in Paris (these correspond to all the stations within more or
less 6 km from Les Halles, for a total of 15 stations). 
The 15 first most populous communes (for year 2000) of the
so-obtained list is in agreement with the ranking published online.

\subsubsection{Pruning the data}

In order to speed up the analysis,
we eliminated the topological nodes of degree 2, taking care of adding up the lengths and travel time of segments.
This reduces the number of links and nodes in the analyzed network, and slightly increases the number of nodes of degree 1 and 3.
We checked that the average travelling speed remains unchanged, 
as well as the number of loops and of connected components (See the next section and Fig.~\ref{fig:simpl_links_nodes}).
The total number of nodes of degree different from 2 (including leaves) is quite low with respect to the total number of nodes or stations, suggesting that most stations are actually of degree 2, i.e. are positioned along a line connecting two major cities
(Fig.~\ref{fig:simpl_links_nodes} a). 

\subsection{General topological measures}

\begin{figure}
\centering
\includegraphics[width=0.5\textwidth]{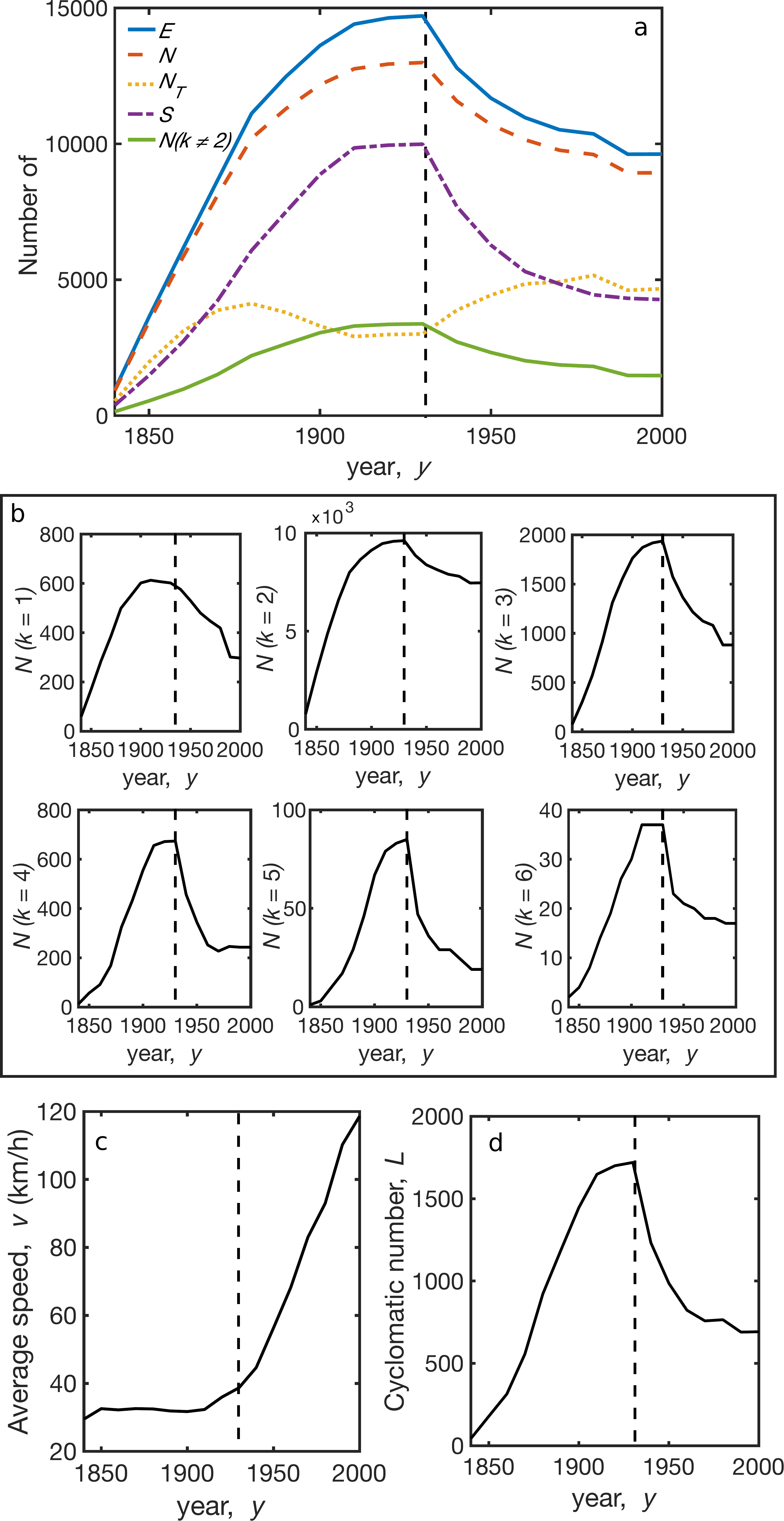}
\caption{\small  Analysis of the pruned network. Evolution versus year of: a) the number of different network's elements (links, nodes, topological nodes, stations, all nodes with degree different from 2); b) the number of nodes of different degree $k$; c) the average travelling speed, d) the number of loops expressed through the cyclomatic number. The dashed vertical line indicates the year 1930.}
\label{fig:simpl_links_nodes}
\end{figure}

\begin{figure}
\centering
\includegraphics[width=0.5\textwidth]{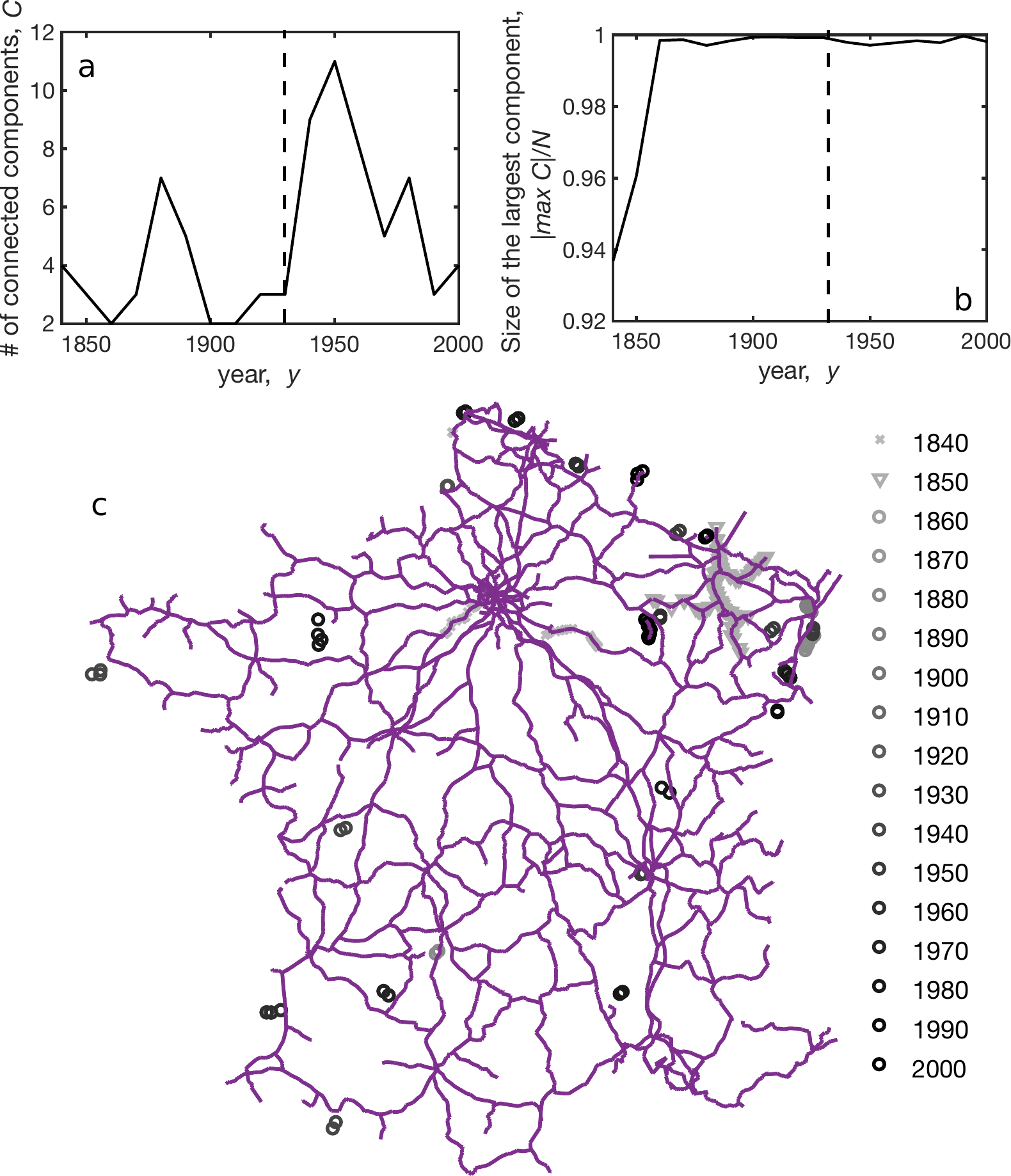}
\caption{\small  Analysis of the network's connected components. Evolution of a) the number of connected components; b) the relative size of the largest connected component. The vertical dashed line corresponds to 1930. c) Spatial plot of the disconnected components over the years superposed to the train network in the year 2000.}
\label{fig:conn_comp}
\end{figure}

The evolution of the number of nodes with a given degree is plotted in 
Fig.~\ref{fig:simpl_links_nodes}b.
Due to the constraint of planarity it is unlikely to observe nodes with a large degree.
Figure~\ref{fig:simpl_links_nodes}c shows the average travelling speed
and Fig.~\ref{fig:simpl_links_nodes}d shows the number of loops, as a function of time.
Loops are simple cycles, i.s. cycles that do not visit
any vertex more than once.  The number of independent cycles in an
undirected graph (called cyclomatic number) is given by $L = E - N + C$,
where $C$ is the number of connected components.  This is also the
minimum number of edges that must be removed from the graph to break
all its cycles, making it into a tree or forest.

Two well-separated phases of growth and shrinking are clearly visible in all graphs.

We also notice that the railway network always has more than one connected component, meaning that some
(small) parts of the network remain isolated [Fig.~\ref{fig:conn_comp}a].
However, the size of the largest connected component is of the order of network size, except in 1840
(Fig.~\ref{fig:conn_comp}b). 
Moreover, the position of the disconnected components does not seem to be strategical for the whole connectivity of the network except in 1840 and possibly in 1850 (Fig.~\ref{fig:conn_comp}c).
These small disconnected components should not substantially affect large-scale quantities such as efficiency and cost, or betweenness centrality, and we therefore discard their existence during our analysis.

\subsection{Cost, efficiency, robustness}

\begin{figure}
\centering
\includegraphics[width=0.45\textwidth]{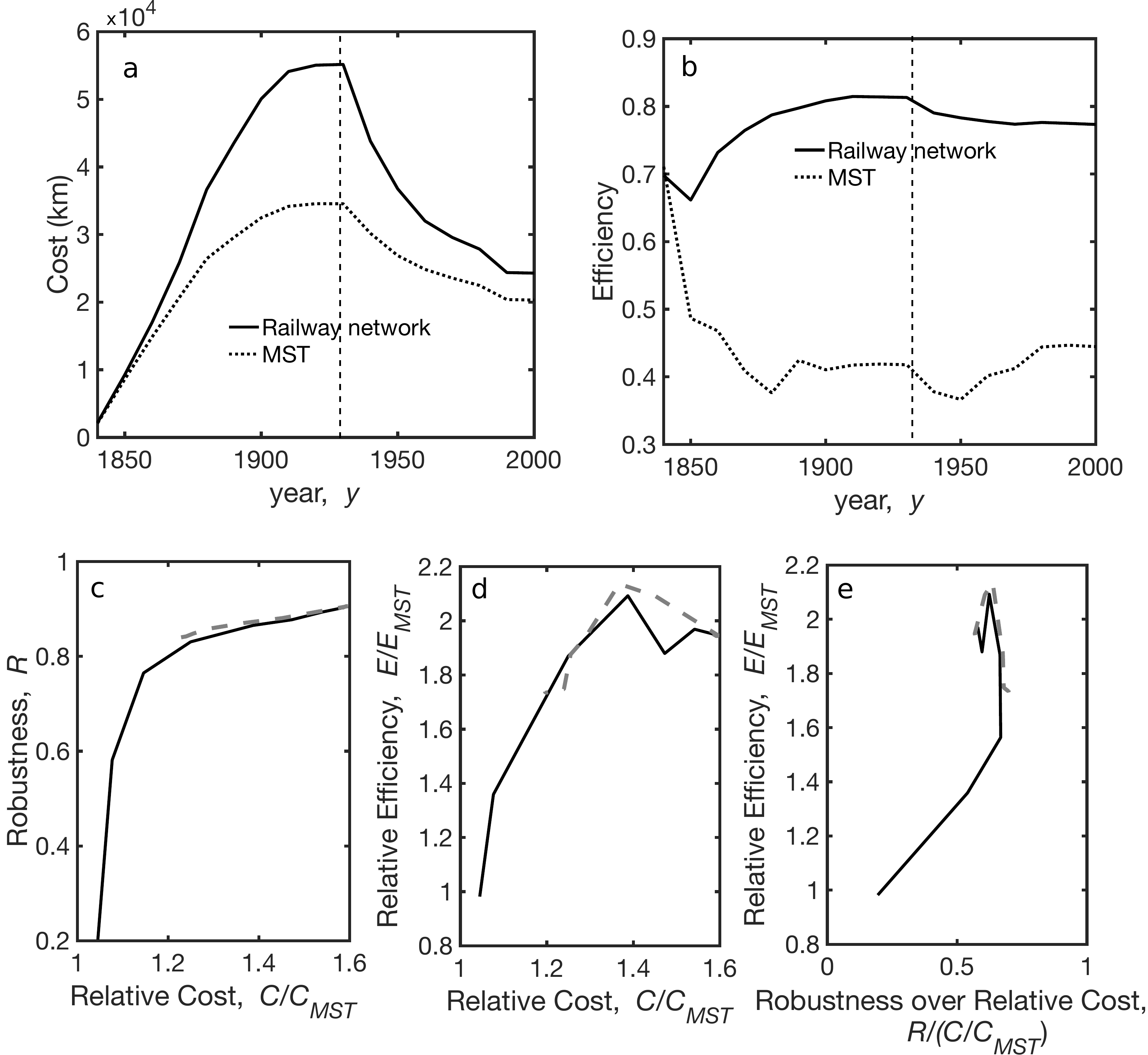}
\caption{\small  Evolution of a) the total cost of the train network; b) the efficiency of the train network compared to the efficiency of the corresponding MST. The dashed vertical line indicates the year 1930. Evolution of the trade-offs between design goals: c) Robustness vs relative cost; d) relative transport efficiency vs relative cost; e) transport efficiency vs network functionality. The evolution before 1930 is the full black line, after is the dashed grey line.}
\label{fig:optimality}
\end{figure}

Figure~\ref{fig:optimality} a and b show the network's cost and efficiency not normalised by the corresponding MST (see the main text for the definitions).
By following the approach taken by Tero et al.~\cite{Tero:2010}, we visualize the evolution of the trade-off between these design goals (normalized by the value taken for the corresponding MST) by plotting these quantities one against each other (Figure~\ref{fig:optimality}c, d, e).
These plots clearly show how the shrinking phase reduced costs while leaving robustness and efficiency almost unchanged.


\subsection{Detour index}

\begin{figure}[h!]
\centering
\includegraphics[width=0.4\textwidth]{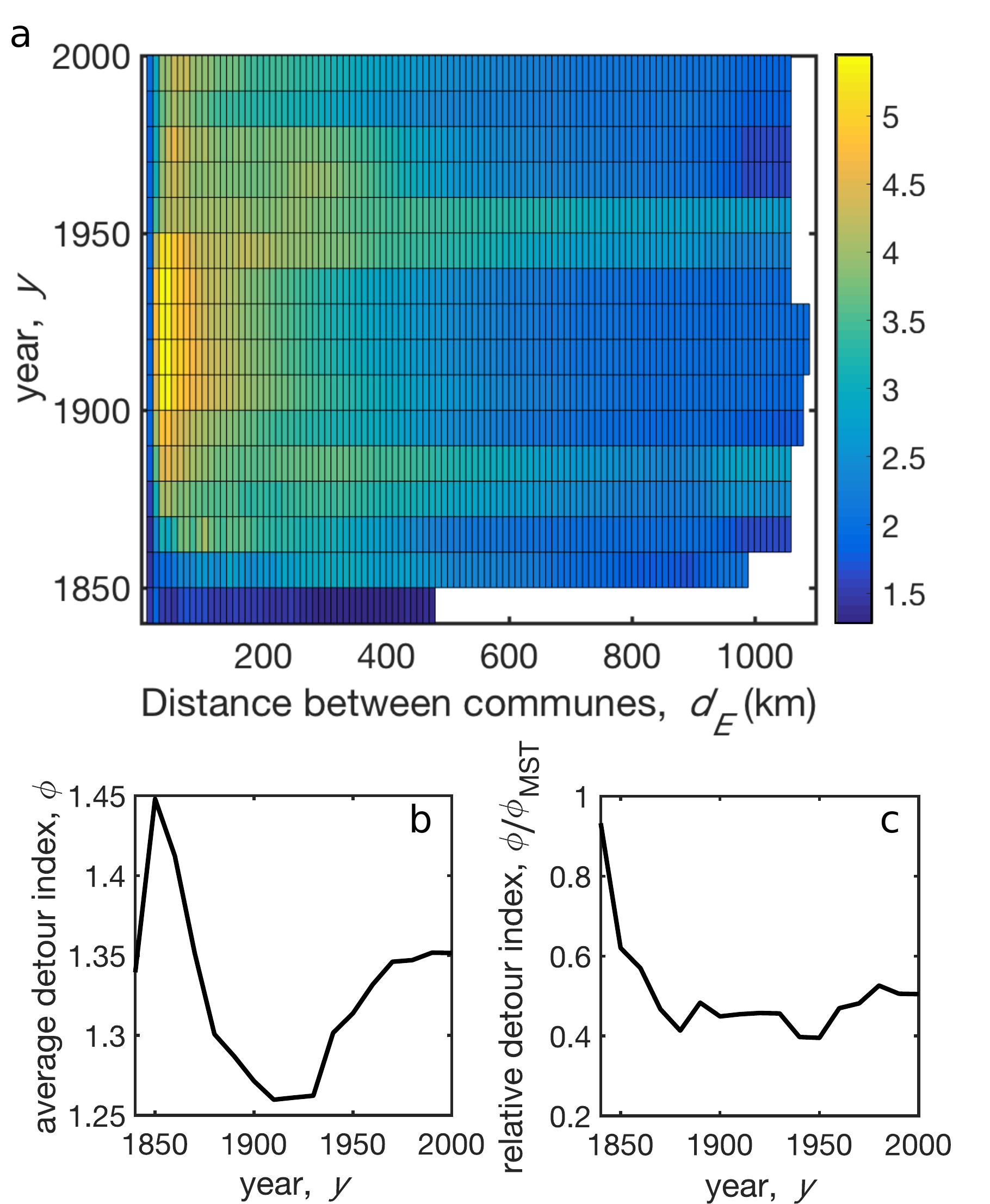}
\caption{\small (a) Detour index over time and at different euclidean
  distances between stations for the MST built on the same set of
  nodes as the current railway network, $\phi_{MST}$. (brighter is high detour index, darker is low detour index). (b) Average detour index for the French railway network, $\phi$ and (c) relative detour index (detour index of the train network divided by the detour index of the corresponding MST) versus time. The vertical dashed line indicates the beginning of the shrinking phase in 1930.}
\label{fig:SIavgdetour}
\end{figure}

While in the main text we widely discussed the behaviour of the detour
index as a function of the distance between stations, $\phi(d)$, for
comparison in figure~\ref{fig:SIavgdetour}a we show the detour index for the
MST built on the same nodes as the actual network. By comparing with
Figure 6 in the main text it is possible to appreciate the evolution
in the optimality of the French railway system and how design goals
have been balanced during both growth and shrinking. Moreover, we show the behaviour of the detour profile averaged over distances and divided by the detour index of the corresponding MST~\ref{fig:SIavgdetour}b and c. 
As the network expands until 1920 we observe a decrease of the average detour index: in 1910, there is on average a $25\%$ difference between the trip on the rail network and the Euclidean distance. After 1920, in the shrinking phase, the average detour index increases due to the removal of narrow gauge lines. However, when compared to the MST, which is the most economical network but known to have a high detour profile, the relative detour profile remains roughly constant, indicating that the efficiency of the network is preserved as the cost is decreased.

\subsection{Population analysis}

\begin{figure}
\centering
\includegraphics[width=0.4\textwidth]{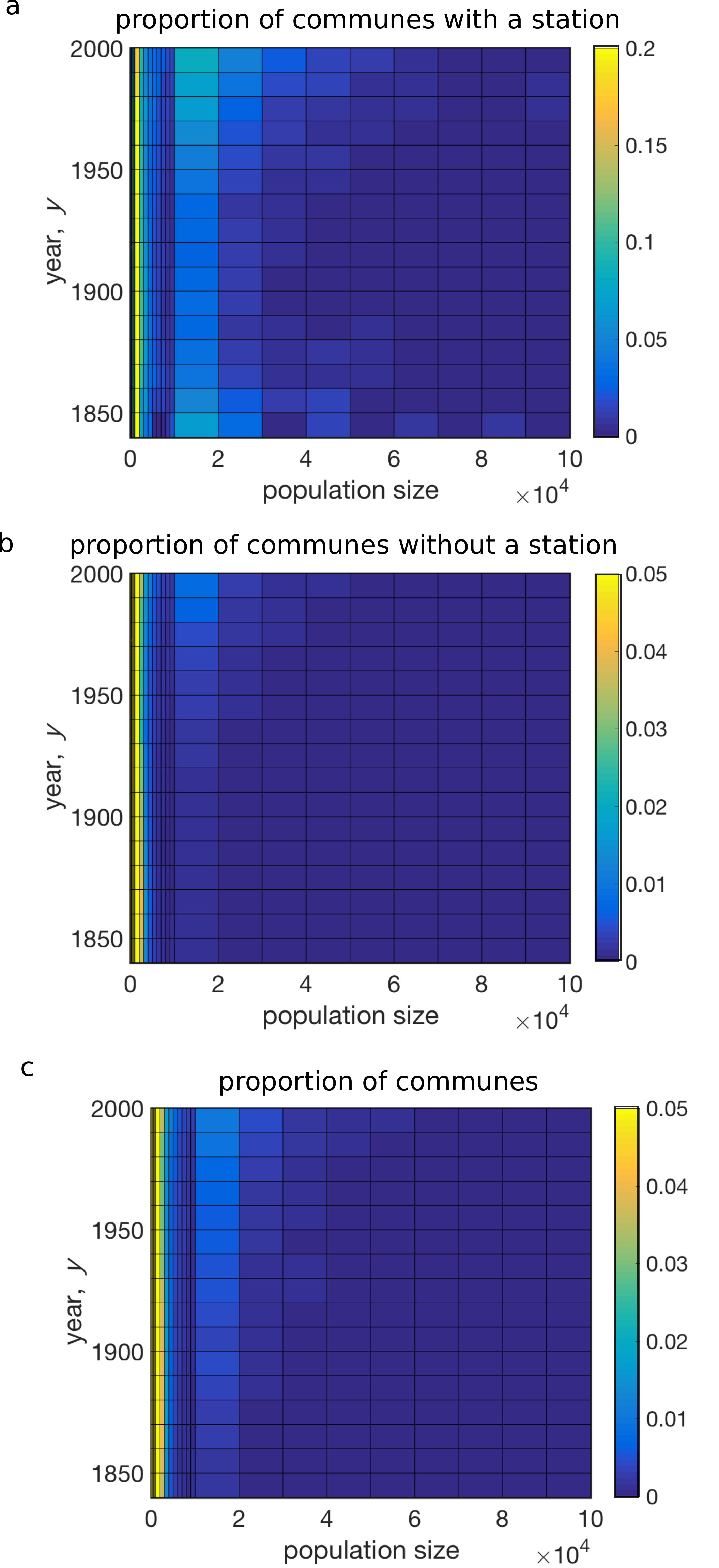}
\caption{\small Evolution of the distribution of communes a) with a station b) without a station c) all communes. The color range for plots b and c is the same, and it has been reduced between 0 and 0.05 to better appreciate the increase in the proportion of larger communes in recent years.}
\label{fig:psd}
\end{figure}

The evolution of the average population for communes (with or without a station)
is reported in the main text.
Here, we complement those plots by showing the full distribution
of population size for communes with a station, without a station, and for all communes considered together
(Fig.~\ref{fig:psd}). From the heath maps it is clear that the largest proportion of communes has a small population, which slightly increased after 1900. Figure~\ref{fig:psd}a also clearly shows that during the capillarization phase between 1890 and 1930 most new stations had been added to smaller communes, and that the shrinking phase was characterised by the elimination of stations in these same small communes.

\begin{figure}
\centering
\includegraphics[width=0.4\textwidth]{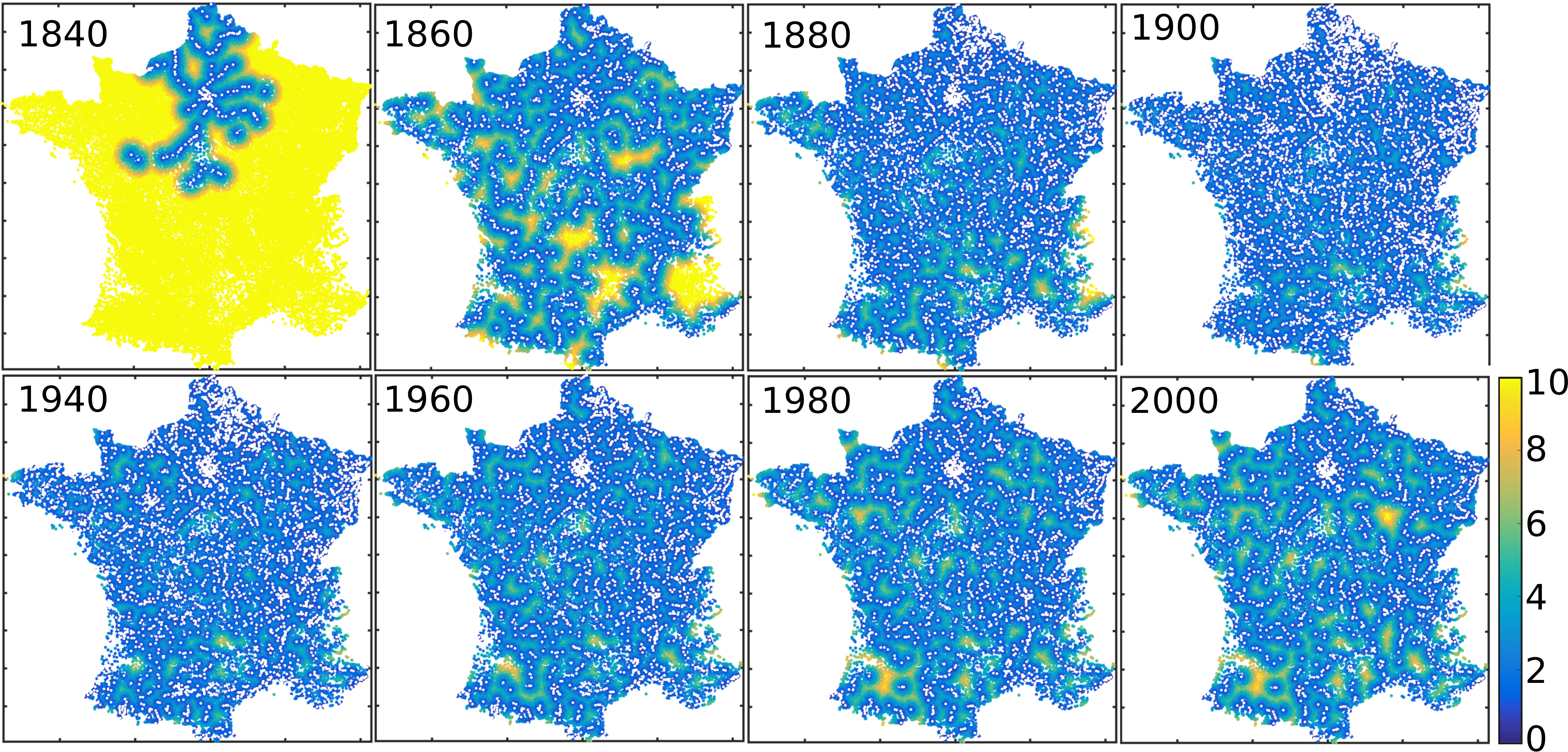}
\caption{\small  Accessibility over time for communes without a station (large circles). Communes with a station are represented by white dots. Notice that the color range is fixed between 0 and 10 hours for all plots for easier comparison.}
\label{fig:Ap}
\end{figure}

\begin{figure}
\centering
\includegraphics[width=0.4\textwidth]{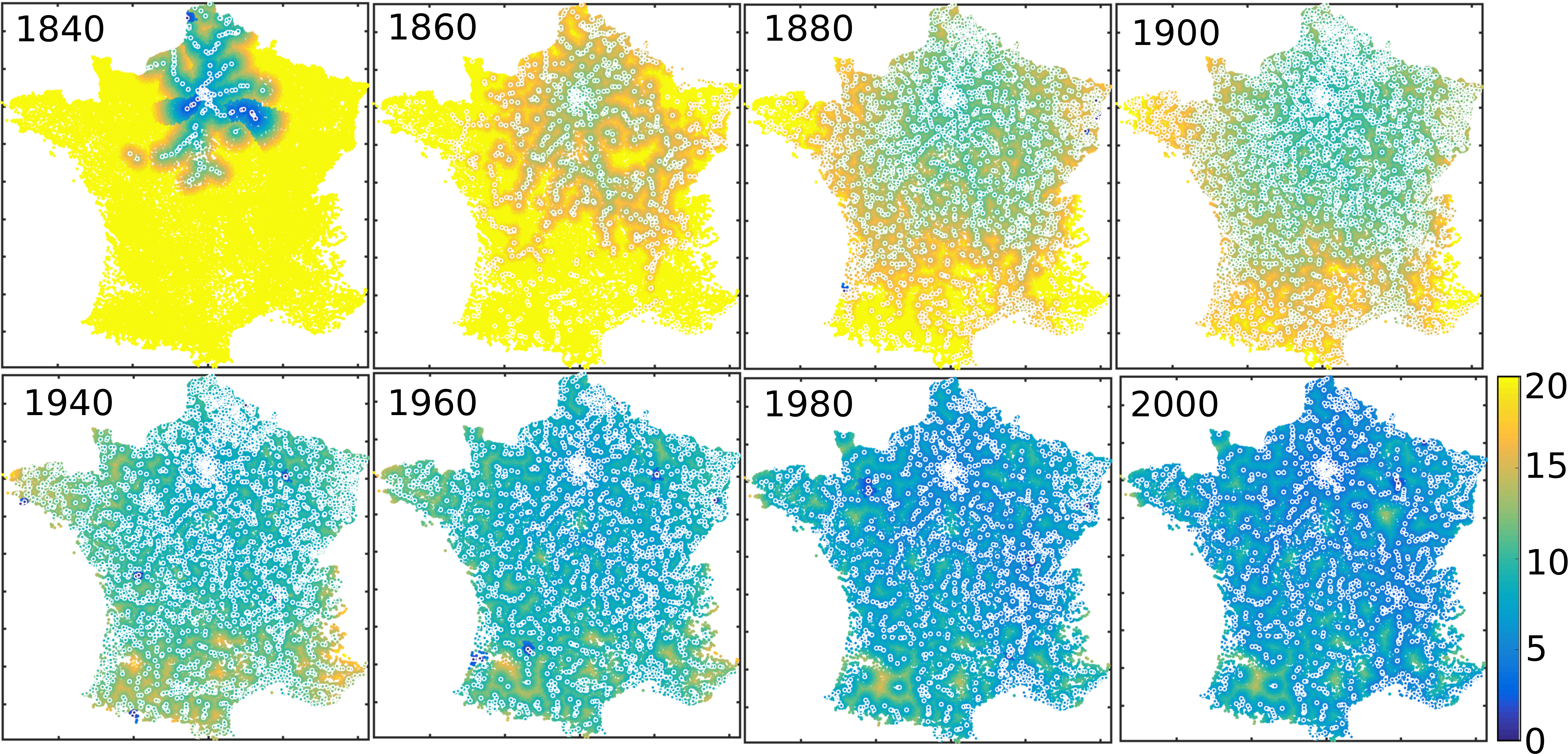}
\caption{\small Generalized accessibility for communes without a station (full circles) and communes with a station (circles with white edge line). The color range is fixed between 0 and 20 hours for all plots.
}
\label{fig:GAp}
\end{figure}

Finally, the local accessibility and generalized accessibility expressed in hours of travel (see main text for the definitions)
are reported as heat maps in Fig.~\ref{fig:Ap} and ~\ref{fig:GAp}, for selected years.

\section{Bibliography}

\begin{thebibliography}{99}

\bibitem{Barrat:2008}
Barrat A, Barthelemy M, Vespignani A (2008) Dynamical Processes on Complex Networks (Cambridge University Press, Cambridge).

\bibitem{Latora:2017} 
Latora V, Nicosia V, Russo G (20170 Complex networks: principles, methods and applications (Cambridge University Press).

\bibitem{Bottinelli:2017}
 Bottinelli A, Louf R, Gherardi M (2017) Balancing building and maintenance costs in growing transport networks. \emph{Physical Review E} 96:032316.

\bibitem{Xie:2011}
Xie F, Levinson D (2011) Evolving transportation networks (Springer Science $\&$ Business Media).

\bibitem{Tero:2010}
Tero A, Takagi S, Saigusa T, Ito K, Bebber DP, Fricker MD, Yumiki K, Kobayashi R,  Nakagaki T (2010) Rules for biologically inspired adaptive network design. \emph{Science} 327:439.

\bibitem{Bottinelli:2015}
Bottinelli A, van Wilgenburg E, Sumpter DJT, Latty T (2015) Local cost minimization in ant transport networks: from small-scale data to large-scale trade-offs. \emph{Journal of The Royal Society Interface} 12:20150780.

\bibitem{Barthelemy:2018}
Barthelemy M (2018) Morphogenesis of Spatial Networks (Springer).

\bibitem{Mimeur:2018}
Mimeur C, Queyroi F, Banos A, Th\'evenin T (2018) Revisiting the structuring effect of transportation infrastructure: an empirical approach with the French Railway Network from 1860 to 1910. \emph{Historical Methods: A Journal of Quantitative and Interdisciplinary History} 51:65.

\bibitem{Moore:2006} 
Moore C, Ghoshal G, Newman MEJ (2006) Exact solutions for models 
of evolving networks with addition and deletion of nodes. \emph{Phys. Rev. E} 74:036121.

\bibitem{Latty:2011}
Latty T, Ramsch K, Ito K, Nakagaki K, Sumpter DJ, Middendorf M, Beekman M (2011) Structure and formation of ant transportation networks. \emph{Journal of The Royal Society Interface} 8:1298.

\bibitem{Perna:2012}
Perna A, Granovskiy B, Garnier S, Nicolis SC, Lab\'edan M, Theraulaz G, 
Fourcassi\'e V, Sumpter DJT (2012)  Individual rules for trail pattern formation in Argentine ants (Linepithema humile). \emph{PLoS Computational Biology} 8:2592.

\bibitem{Ma:2013}
Ma Q, Johansson A, Tero A, Nakagaki T, Sumpter DJT (2013) Current-reinforced random walks for constructing transport networks. \emph{Journal of The Royal Society Interface} 10:20120864.

\bibitem{Grauwin:2009}
Grauwin S, Bertin E, Lemoy R, Jensen P (2009) Competition between collective and individual dynamics. \emph{Proc. NAtl. Acad. Sci. (USA)} 106:20622.

\bibitem{Bouchaud:2013}
Bouchaud JP (2013)  Crises and collective socio-economic phenomena: simple models and challenges. \emph{Journal of Statistical Physics} 151:567.

\bibitem{Bertolini:2007}
Bertolini L (2007) Evolutionary urban transportation planning: an exploration. \emph{Environment and Planning A} 39:1998.

\bibitem{Thevenin:2016}
Thevenin T, Mimeur C, Schwartz R, Sapet L (2016) Measuring one century of railway accessibility and population change in France. A historical GIS approach. \emph{Journal of Transport Geography} 56.

\bibitem{Clark:1991}
Clark J, Holton D (1991) A first look at graph theory (World Scientific).

\bibitem{Latora:2007}
Latora V, Marchiori M (2007) A measure of centrality based on network efficiency. \emph{New Journal of Physics} 9:188.

\bibitem{Graham:1985}
Graham RL, Hell P (1985) On the history of the minimum spanning tree problem. \emph{Annals of the History of Computing} 7:43.

\bibitem{Cardillo:2006}
Cardillo A, Scellato S, Latora V, Porta S (2006) Structural properties of planar graphs of urban street patterns. \emph{Physical Review E} 73:066107.

\bibitem{Latora:2001}
Latora V, Marchiori M (2001) Efficient behavior of small-world networks. \emph{Physical Review Letters} 87:198701.

\bibitem{Cook:2014}
Cook Z, Franks DW, Robinson EJ (2014) Efficiency and robustness of ant colony transportation networks. \emph{Behavioral Ecology and Sociobiology} 68:509.

\bibitem{Aldous:2010}
Aldous DJ, Shun J (2010) Connected spatial networks over random points and a route-length statistic. \emph{Statistical Science} 25:275.

\bibitem{Gastner:2006}
Gastner M, Newman M (2006) Optimal design of spatial distribution networks. \emph{Phys. Rev. E} 74:016117.

\bibitem{Bebber:2007}
Bebber DP, Hynes J, Darrah PR, Boddy L, Fricker MD (2007) Biological solutions to transport network design. \emph{Proceedings of the Royal Society B: Biological Sciences} 274:2307.


\bibitem{Handy:1997}
Handy SL, Niemeier DA (1997) Measuring accessibility: an exploration of issues and alternatives. \emph{Environment and planning A} 29:1175.

\bibitem{Geurs:2004} 
Geurs KT, Van Wee B (2004) Accessibility evaluation of land-use and transport strategies: review and research directions. \emph{Journal of Transport geography} 12:127.


\end{thebibliography}
\end{document}